\documentclass[conference,10pt]{IEEEtran}
\IEEEoverridecommandlockouts
\usepackage{cite}
\usepackage{amsmath,amssymb,amsfonts}
\usepackage{algorithmic}
\usepackage{graphicx}
\usepackage{textcomp}
\usepackage{xcolor}
\usepackage{bm}
\usepackage{algorithm}
\usepackage{algorithmic}
\usepackage{soul}
\usepackage[top=0.65in, bottom=1.22in, left=0.61in, right=0.62in]{geometry}

\def\BibTeX{{\rm B\kern-.05em{\sc i\kern-.025em b}\kern-.08em
    T\kern-.1667em\lower.7ex\hbox{E}\kern-.125emX}}
\begin{document}

\title{\huge
	 Near-Field Sparse Channel Estimation for Extremely Large-Scale RIS-Aided Wireless Communications\\
}

\author{{\large Zixing~Tang\textsuperscript{1,*},~Yuanbin~Chen\textsuperscript{1,*},~Ying~Wang\textsuperscript{1},~Tianqi~Mao\textsuperscript{2},~Qingqing~Wu\textsuperscript{3},}\\{\large Marco~Di~Renzo\textsuperscript{4},~Lajos~Hanzo\textsuperscript{5} }\\
	{\normalsize \textsuperscript{1}State Key Laboratory of Networking and Switching Technology,}\\ {\normalsize Beijing University of Posts and Telecommunications, Beijing 100876, China}\\
	{\normalsize \textsuperscript{2}School of Information and Electronics and the Advanced Research Institute of Multidisciplinary Science,}\\ {\normalsize Beijing Institute of Technology, Beijing 100081, China}\\
	{\normalsize \textsuperscript{3}Department of Electronic Engineering, Shanghai Jiao Tong University, Shanghai 200240, China}\\
	{\normalsize \textsuperscript{4}Universit\'{e} Paris-Saclay, CNRS, CentraleSup\'{e}lec, Laboratoire des Signaux et Syst\`{e}mes,} \\{\normalsize 3 Rue Joliot-Curie, 91192 Gif-sur-Yvette, France}\\
	{\normalsize \textsuperscript{5}University of Southampton, Southampton SO17 1BJ, U.K.}\\
	
\vspace{-5mm}
}

\maketitle

\renewcommand{\thefootnote}{\fnsymbol{footnote}}
\footnotetext[1]{Zixing Tang and Yuanbin Chen contributed equally to this work.}
\renewcommand{\thefootnote}{\arabic{footnote}}
\setcounter{footnote}{0}

\begin{abstract}
A significant increase in the number of reconfigurable intelligent surface (RIS) elements results in a spherical wavefront in the near field of extremely large-scale RIS (XL-RIS). Although the channel matrix of the cascaded two-hop link may become sparse in the polar-domain representation, their accurate estimation of these polar-domain parameters cannot be readily guaranteed. To tackle this challenge, we exploit the sparsity inherent in the cascaded channel. To elaborate, we first estimate the significant path-angles and distances corresponding to the common paths between the BS and the XL-RIS. Then, the individual path parameters associated with different users are recovered. This results in a two-stage channel estimation scheme, in which distinct learning-based networks are used for channel training at each stage. More explicitly, in stage I, a denoising convolutional neural network (DnCNN) is employed for treating the  grid mismatches as noise to determine the true grid index of the angles and distances. By contrast, an iterative shrinkage thresholding algorithm (ISTA) based network is proposed for adaptively adjusting the column coherence of the dictionary matrix in stage II. Finally, our simulation results demonstrate that the proposed two-stage learning-based channel estimation outperforms the state-of-the-art benchmarks.

\end{abstract}


\section{Introduction}
Recent academic and industrial interest in reconfigurable intelligent surfaces (RISs) has been fueled by its potential to improve the spectral and/or energy efficiency of communication systems \cite{Rose,Ruiqi}. However, they also pose challenges. For example, when the communication link between the base station (BS) and the user is adequate, their performance gain becomes negligible. This is primarily due to the double fading experienced by the transmitted signal in the two-hop RIS-aided system, which is determined by the multiplicative cascaded channel gain model \cite{chen-vtm}. To mitigate the deleterious effects of multiplicative twin-hop channel gain, the number of RIS elements may be increased, thus resulting in the concept of the extremely large-scale RIS (XL-RIS). Nonetheless, further challenges arise. Firstly, a significant increase in the number of RIS elements exacerbates the estimation of cascaded channels, since the number of cascaded channel coefficients to be determined will be given by the number of RIS elements. Secondly, the resultant Rayleigh distance expansion renders the commonly used planar-wavefront assumption invalid due to the spherical wavefront shape experienced in the near-field region.

To mitigate the excessive complexity and overhead of the RIS channel estimation, significant efforts have been devoted to exploiting the potential sparsities of the cascaded channel to achieve pilot overhead savings \cite{RIS-CE-1,RIS-CE-3,chen-twc3, guo-tcom, JSTSP, tvt}. Specifically, due to the limited number of significant paths induced by the sparsity of scatterers in the environment, the cascaded channel exhibits sparsity in the angular domain. These sparse channels may be efficiently estimated by employing compressive sensing (CS) based techniques, such as the popular orthogonal matching pursuit (OMP) \cite{RIS-CE-1,RIS-CE-3} and the variational Bayesian Inference (VBI) algorithms \cite{chen-twc3, guo-tcom}. However, these investigations are predicated on the assumption of a planar wavefront in the far-field region, since the Rayleigh distance is typically a few meters and may be negligible in practice. However, in the XL-RIS regime relying on an extremely large number of elements, the near-field effect can no longer be disregarded due to the resultant energy spread effect. More explicitly, the transmit power may spread across multiple directions, instead of being concentrated in the direction of the desired destination \cite{XLM-1}. To tackle this issue, a polar-domain representation that simultaneously accounts for both the angle and distance information has been proposed in \cite{XLM-1}.

However, there are two major issues if polar-domain sparsity is assumed for the cascaded channel in the XL-RIS regime. Firstly, in the presence of the polar-domain representation, the nonzero elements of the cascaded channel matrix (whose physical meanings correspond to the angles and distances associated with significant paths) may not fall precisely on the uniform grids prescribed by the dictionary matrix, leading to grid mismatches. As such, the power of nonzero elements in the cascaded channel matrix may leak to other elements, hence resulting in a power leakage effect. Secondly, due to the inherent column coherence of the dictionary matrix, the single power peak corresponding to the one and only significant path may drift to different grid indices, which is referred to as the ``power drift" phenomenon. Therefore, given this pair of challenges, a tailor-made channel estimation scheme is required for the XL-RIS regime .

\addtolength{\topmargin}{-0.1in}

Inspired by the dual-structured sparsity of the cascaded channel presented in \cite{chen-twc3}, we propose a two-stage learning-based channel estimation approach, with the aim of embracing the challenges inherent in near-field XL-RIS aided wireless communications. Specifically, upon assuming polar-domain sparsity for the near-field cascaded channel matrix, the resultant sparse cascaded channel matrix shares the same nonzero rows, but it has different nonzero columns for each user. To this end, the goal of our two-stage channel estimation is to estimate the significant angles and distances corresponding to the common paths of all users (contained in the nonzero rows) spanning from the BS to the XL-RIS in its stage I. By contrast, in stage~II, we recover the distances and angles at the RIS, as well as the complex gains associated with each user. More explicitly, in stage I, a denoising convolutional neural network (DnCNN) is employed, which treats grid mismatches as noise, when aiming for retrieving the true grid index of the angles and distances corresponding to the common paths. This effectively mitigates the power leakage effect induced by the grid mismatch. For stage II, we propose a regime termed as the iterative shrinkage thresholding algorithm (ISTA) based network for adaptively adjusting the column coherence of the dictionary matrix to mitigate the power drift effect. We will also demonstrate that the proposed two-stage channel estimation scheme has the potential to significantly reduce the gridding complexity and the estimation overhead. Finally, our simulation results demonstrate that the proposed approach achieves accurate estimation of the near-field cascaded channel in XL-RIS aided wireless systems.

\section{System and channel model}
\addtolength{\topmargin}{-0.05in}
\subsection{System Model}

\begin{figure}[t]
\centerline{\includegraphics[width=0.48\textwidth]{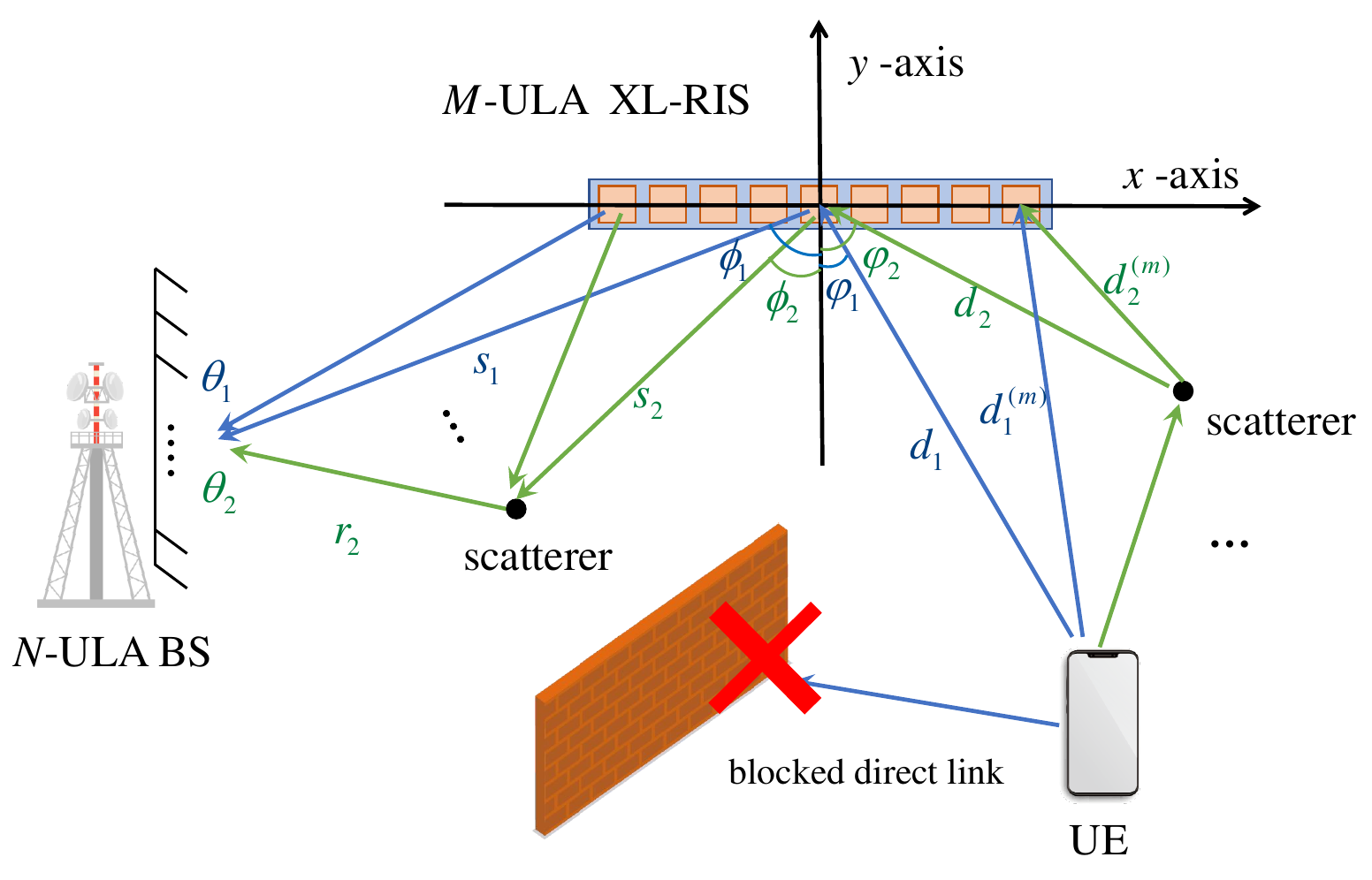}}
\caption{XL-RIS system in the near field.}
\label{system_model}
\end{figure}

We consider the narrow-band XL-RIS uplink communication system, where $K$ single-antenna users are served by an $N$-antenna uniform linear array (ULA) assisted BS. The XL-RIS is equipped with $M$ passive reflecting elements in the form of a ULA. As illustrated in Fig.~\ref{system_model}, the direct link is blocked by obstacles. Therefore, we only focus on the uplink channel estimation of the XL-RIS assisted link.

Let us consider an arbitrary user and assume that a coherence block contains $T$ time slots. Let $\mathbf{H} \in \mathbb{C}^{N \times M}$ denote the near-field channel spanning from the XL-RIS to the BS and $\mathbf{h} \in \mathbb{C}^{M \times 1}$ denote the near-field channel impinging from the user upon the XL-RIS. Moreover, denote the phase shift vector in time slot $t$ by $\mathbf{e}_t \in \mathbb{C}^{M \times 1}$. In the first $\tau$ slots, the user transmits an orthogonal pilot sequence $\mathbf{s} \in \mathbb{C}^{1 \times K}$, which satisfies $\mathbb{E}\left[ \mathbf{s}\mathbf{s}^H\right] =1$. Specifically, the signal received at the BS in slot $t, 1 \leq t \leq \tau$ is given by
\begin{equation}
\mathbf{Y}(t) = \mathbf{H} \text{diag}(\mathbf{e}_t) \mathbf{h} \sqrt{p} \mathbf{s} + \mathbf{N}(t),
\end{equation}
where $p$ represents the transmit power of the user, while $\mathbf{N}(t) \in \mathbb{C}^{M \times K}$ is the additive white Gaussian noise (AWGN) following the distribution $\text{vec}(\mathbf{N}(t)) \sim\mathcal{CN}(0, \sigma^2 \mathbf{I})$. Next, we right-multiply the received signal by the pilot sequence, yielding
\begin{equation}
\begin{aligned}
\mathbf{Y}(t)\mathbf{s}^H &= \sqrt{p}\mathbf{H} \text{diag}(\mathbf{e}_t) \mathbf{h} + \mathbf{N}(t)\mathbf{s}^H \\ &= \sqrt{p}\mathbf{H} \text{diag}(\mathbf{h})\mathbf{e}_t  + \mathbf{N}(t)\mathbf{s}^H.
\end{aligned}
\end{equation}
Furthermore, let us denote the cascaded channel by $\mathbf{G} = \mathbf{H}\text{diag}(\mathbf{h})\in \mathbb{C}^{N \times M}$. Upon stacking $\tau$ signals, the overall received signal matrix $\mathbf{Y} = [\mathbf{Y}(1)\mathbf{s}^H, \dots, \mathbf{Y}(\tau)\mathbf{s}^H] \in \mathbb{C}^{N \times \tau}$ is expressed as
\begin{equation}\label{received}
\mathbf{Y} = \sqrt{p}\mathbf{G}\mathbf{E} + \mathbf{N},
\end{equation}
where $\mathbf{E} = [\mathbf{e}_1, \dots, \mathbf{e}_\tau]$ and $\mathbf{N} = [\mathbf{N}(1)\mathbf{s}^H, \dots, \mathbf{N}(\tau)\mathbf{s}^H]$.


\subsection{Near-field Cascaded Channel Sparsity Model}

It is assumed that there are $L_{BR}$ propagation paths between the XL-RIS and the BS, as well as $L_{RU}$ paths between the user and the XL-RIS. The near-field channel between the XL-RIS and the BS, as well as the one between the user and the XL-RIS are modeled as \cite{near_0}
\begin{equation}\label{tmp-4}
\mathbf{H} = \mathbf{A}_N(\bm{\theta}, \bm{r}) \text{diag}(\bm{\rho}) \mathbf{A}_M^H(\bm{\phi}, \bm{s}),
\end{equation}
\begin{equation}
\mathbf{h} = \mathbf{A}_M(\bm{\varphi}, \bm{d}) \bm{\beta},
\end{equation}
where $\mathbf{A}_N(\bm{\theta}, \bm{r}) \in \mathbb{C}^{N \times L_{BR}}$, $\mathbf{A}_M(\bm{\phi}, \bm{s}) \in\mathbb{C}^{M \times L_{BR}}$ and $\mathbf{A}_M(\bm{\varphi}, \bm{d}) \in\mathbb{C}^{M \times L_{RU}}$ represent the near-field steering matrices having angles of arrival/departure (AoAs/AoDs) and distances. Furthermore, $\bm{\rho} \in \mathbb{C}^{L_{BR} \times 1}$ and $\bm{\beta} \in \mathbb{C}^{L_{RU} \times 1}$ represent the complex gains of the paths between the XL-RIS and the BS, as well as those between the user and the XL-RIS. As illustrated in Fig.~\ref{system_model}, $\bm{\theta}$ denotes the AoAs at the BS, while $\bm{\phi}$ and $\bm{\varphi}$ are the AoDs and AoAs at the XL-RIS, respectively. Still referring to (\ref{tmp-4}), $\bm{r}$ represents the distances between  the BS and the scatterers (or XL-RIS), while $\bm{s}$ and $\bm{d}$ represent the distances between the scatterers (or user/BS) and the XL-RIS.

Specifically, the near-field steering matrices are given by
\begin{equation}\label{steering_1}
\mathbf{A}_N(\bm{\theta}, \bm{r})= \left[\mathbf{a}_N(\theta_1, r_1), \dots, \mathbf{a}_N(\theta_{L_{BR}}, r_{L_{BR}})\right],
\end{equation}
\begin{equation}\label{steering_2}
\mathbf{A}_M(\bm{\phi}, \bm{s})= \left[\mathbf{a}_M(\phi_1, s_1), \dots, \mathbf{a}_M(\phi_{L_{BR}}, s_{L_{BR}})\right],
\end{equation}
\begin{equation}\label{steering_3}
\mathbf{A}_M(\bm{\varphi}, \bm{d})= \left[\mathbf{a}_M(\varphi_1, d_1), \dots, \mathbf{a}_M(\varphi_{L_{RU}}, d_{L_{RU}})\right],
\end{equation}
where $\mathbf{a}_X(\cdot) \in \mathbb{C}^{X \times 1}$ represents the near-field array steering vector of a propagation path. Taking $\mathbf{a}_M(\varphi_l, d_l)$ as an example, it can be approximated using the Fresnel approximation of the spherical wavefront model by its angle and distance as~\cite{near_1}
\begin{equation}\label{nf_steer}
\mathbf{a}_M(\varphi_l, d_l) = \frac{1}{\sqrt{M}}\left[\dots, e^{-jk(d_l^{(m)}-d_l)}, \dots\right]^T,
\end{equation}
\begin{equation}
d_l^{(m)} \approx d_l - m\delta\sin{\varphi} + \frac{m^2\delta^2\cos^2{\varphi}}{2d_l},
\end{equation}
where $d_l^{(m)}$ represents the distance between the $m$th reflecting element and the $l$th scatterer, and $d_l = d_l^{(0)}$. $\delta$ is the reflecting element spacing, while $k=\frac{2\pi f_c}{c}$ denotes the wavenumber at the central carrier $f_c$. Moreover, $m$ satisfies
$ -\left\lceil\frac{M-1}{2}\right\rceil \leq m \leq\left\lfloor\frac{M-1}{2}\right\rfloor, m \in \mathbb{Z} $.

According to \cite{XLM-1}, the near-field steering matrices exhibit beneficial sparsities in the presence of the polar-domain representation. Let us denote the overcomplete dictionary matrices in the polar domain by $\mathbf{F}_N \in \mathbb{C}^{N \times N_G}$ and $\mathbf{F}_M \in \mathbb{C}^{M \times M_G}$, respectively, with $N_G$ and $M_G$ indicating the number of grids. The steering matrices in (\ref{steering_1})-(\ref{steering_3}) can be recast as $ \mathbf{A}_N(\bm{\theta}, \bm{r}) = \mathbf{F}_N \mathbf{X}_N $, $ \mathbf{A}_M(\bm{\phi}, \bm{s}) = \mathbf{F}_M \mathbf{X}_M^{BR} $, and $ \mathbf{A}_M(\bm{\varphi}, \bm{d}) = \mathbf{F}_M \mathbf{X}_M^{RU} $, respectively,
where $\mathbf{X}_N \in\mathbb{C}^{N_G \times L_{BR}}$, $\mathbf{X}_M^{BR} \in\mathbb{C}^{M_G \times L_{BR}}$ and $\mathbf{X}_M^{RU}\in\mathbb{C}^{M_G \times L_{RU}}$ are sparse matrices. Explicitly, each column vector has exactly one nonzero element. Hence, the cascaded channel $\mathbf{G}$ can be characterized by the sparse matrix $\mathbf{\Lambda} \in \mathbb{C}^{N_G \times M_G^\prime}$, yielding
\begin{equation}\label{G_sparse}
\mathbf{G} = \mathbf{F}_N \mathbf{\Lambda} \tilde{\mathbf{F}}_M^H,
\end{equation}
where $\mathbf{\Lambda}$ has $L_{BR}L_{RU}$ nonzero elements corresponding to the cascaded complex gains $\left\{\rho_l \beta_p\right\}$. It can be formally shown that there are only $M_G^\prime$ distinct column vectors in $\mathbf{F}_M \circ \mathbf{F}_M^* \in \mathbf{C}^{M \times M_G^2}$, where $\circ$ represents the transposed Khatri-Rao product. Therefore, $\tilde{\mathbf{F}}_M \in \mathbf{C}^{M \times M_G^\prime}$ is the combination of $M_G^\prime$ distinct column vectors. Accordingly, the gridding complexity order for a single user is proportional to $\mathcal{O}(N_G M_G)$, with their typical values being $N_G = N$ and $M_G = M$, respectively. Such a high-dimensional input renders channel estimation inefficient.

Again, based on (\ref{G_sparse}), the cascaded channel exhibits dual-structured sparsity of \cite{chen-twc3}, which is a unique property in RIS-aided systems. This property inspires us to separately estimate the common parameters at the BS corresponding to the rows and the parameters at the XL-RIS corresponding to the columns at a reduced gridding complexity.
\addtolength{\topmargin}{-0.05in}

\begin{figure}[t]
	\centerline{\includegraphics[width=0.45\textwidth]{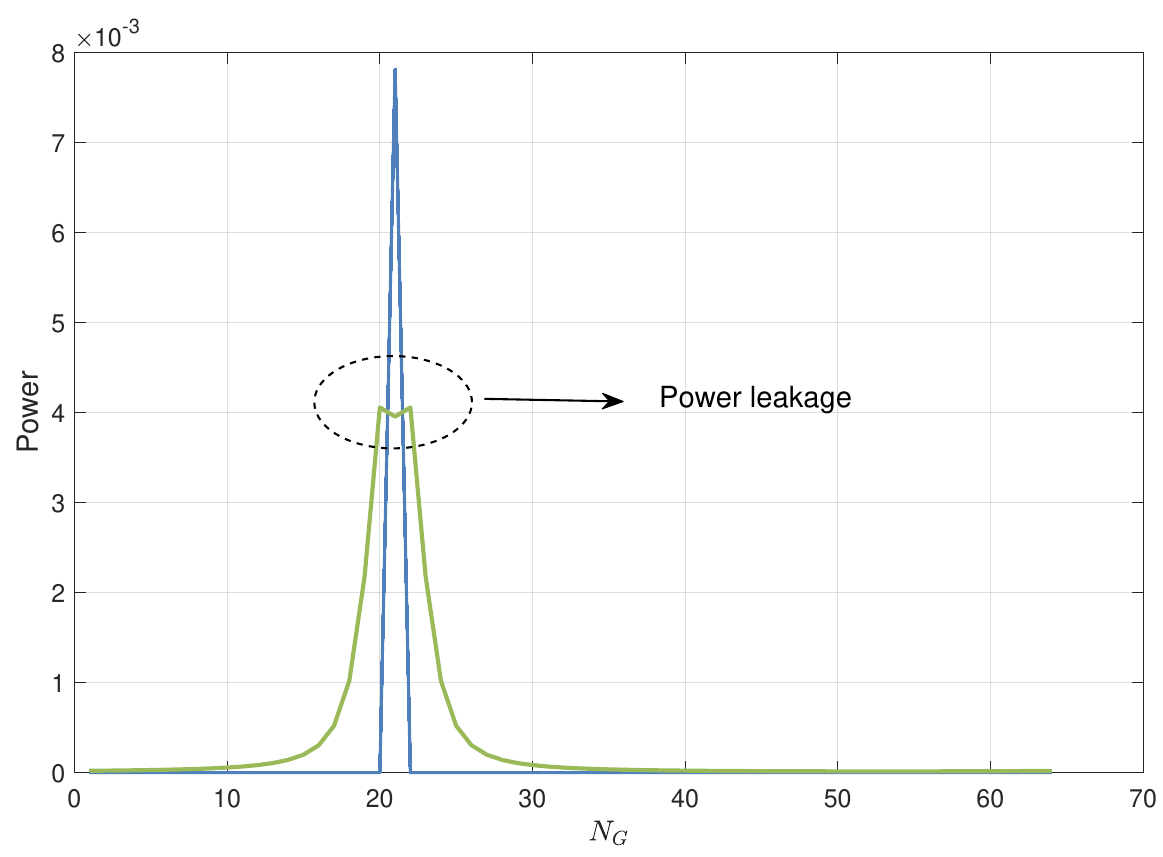}}
	\caption{An example of the row sparsity of $\mathbf{Y}_{Polar}$, where $L_{BR}=1$ and $N=64$.}
	\label{power_leakage_1}
\end{figure}

\section{Two-stage channel estimation}
In this section, we aim for reducing the multiplicative gridding complexity order of $\mathcal{O}(N_G M_G)$, to its additive counterpart, namely to $\mathcal{O}(N_G + M_G)$, by conceiving a two-stage procedure. This beneficially reduces the estimation complexity. Given that all users share the common channel spanning from the XL-RIS to the BS, we decompose the common parameters at the BS from the user-specific parameters at the XL-RIS. Firstly, both the arrival steering matrix corresponding to the common AoAs $\bm{\theta}$ and the distances $\bm{r}$ at the BS are estimated by exploiting the row sparsity of the received signal. Next, given the knowledge of the estimated arrival steering matrix, the estimation of both the cascaded matrix corresponding to the individual AoDs/AoAs $\bm{\phi}, \bm{\varphi}$ and the distances $\bm{s}, \bm{d}$ at the XL-RIS as well as the cascaded complex gains is formulated as a CS problem exploiting the column sparsity. This formulates the proposed two-stage channel estimation procedure. A pair of learning-based networks are harnessed. On the one hand, this approach exhibits robustness to grid-modeling mismatch and achieves faster inference speed than traditional iterative algorithms. On the other hand, the learning-based approach performs well in encoding and decoding the signal, where the substitution of neural networks for the sensing matrix is beneficial in order to make up for the shortcomings of the dictionary matrix in the polar domain~\cite{learning-1,JiangXie}.

\subsection{Stage I: Common BS Steering Matrix Estimation}

Based on (\ref{G_sparse}), we observe that $\mathbf{Y}_{Polar} = \mathbf{F}_N^H \mathbf{Y} \in \mathbb{C}^{N_G \times \tau}$ is a row-sparse and column-full matrix. The indices of nonzero rows are related to the arrival angles and distances at the BS. Accordingly, the estimated arrival steering matrix $\hat{\mathbf{A}}_N(\bm{\theta}, \bm{r})$ can be constructed.

However, the angles $\bm{\theta}$ and distances $\bm{r}$ are continuous values, which may not fall on the discrete grids of the dictionary matrix $\mathbf{F}_N$. Fig.~\ref{power_leakage_1} is an example of the row-sparse structure of $\mathbf{Y}_{Polar} $ and the $y$-axis represents the power of each row of $\mathbf{Y}_{Polar}$. It can be seen that the grid mismatch in distance leads a power peak to spread, which is again referred to as power leakage in this paper. To avoid the resultant performance degradation, we propose a sparse row recovery algorithm based on the DnCNN~\cite{learning-1}.

Briefly, the DnCNN excels at learning the residual noise and separating the noise from a noisy image by harnessing feed-forward convolutional neural networks (CNNs). The image denoising is treated as a plain learning problem, implying that for a propagation path component, both the other path components and the AWGN are treated as residual noises contaminating the desired image (i.e., the path of interest). The process of estimating the steering vector for each propagation path is equivalent to the associated noise removal. The DnCNN-based recovery scheme is capable of beneficially harnessing the leaked power, while removing the noise.

\begin{figure}[t]
	\centerline{\includegraphics[width=0.46\textwidth]{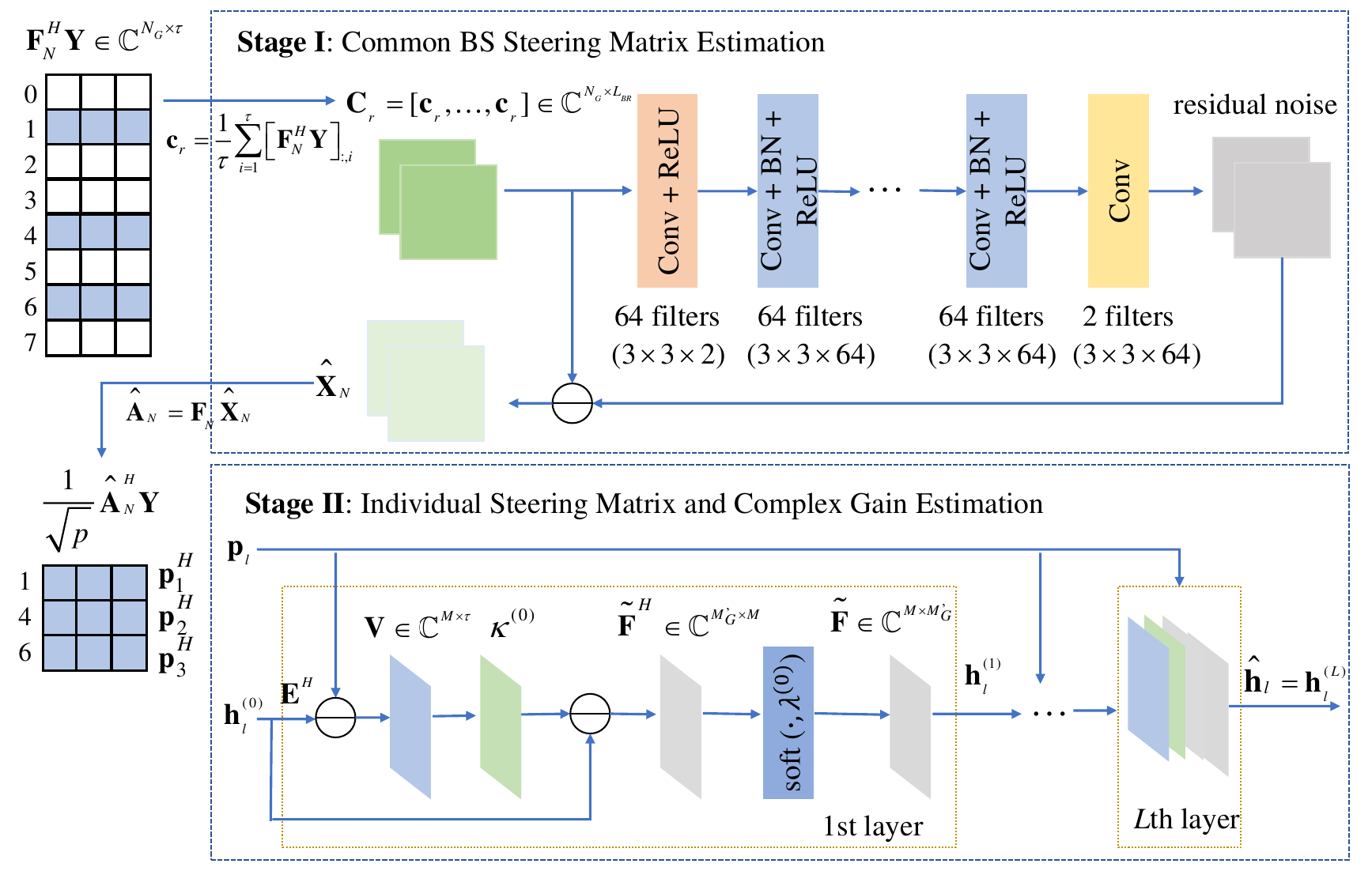}}
	\caption{Cascaded channel estimation scheme.}
	\label{al_model}
\end{figure}

As shown in Fig. \ref{al_model}, the DnCNN consists of $L_c$ convolutional layers. The first convolutional layer is followed by a rectifier linear unit (ReLU). The succeeding $L_c-2$ convolutional layers are followed by batch normalization (BN) and a ReLU. Due to the feed-forward layer's inability to process complex numbers, we split the real and imaginary parts of the input and output into two channels. Specifically, as only the row features have to be emphasized, we sum the column vectors of $\mathbf{Y}_{Polar}$ and define the column vector $\mathbf{c}_r \in \mathbb{C}^{N_G \times 1}$ as
\begin{equation}
\mathbf{c}_r = \frac{1}{\tau}\sum_{i=1}^{\tau}{\left[\mathbf{F}_N^H \mathbf{Y}\right]_{:,i}},
\end{equation}
where $\left[\cdot\right]_{:,i}$ represents the $i$th column of the matrix. To ensure that the input dimension matches the output dimension, we create $L_{BR}$ identical column vectors as the input matrix $\mathbf{C}_r = \left[\mathbf{c}_r, \dots, \mathbf{c}_r\right]\in \mathbb{C}^{N_G \times L_{BR}}$. Let us denote the parameters of the DnCNN by $\bm{\varpi}_1$ and its output by $\mathcal{R}(\mathbf{C}_r;\mathbf{\bm{\varpi}}_1) \in \mathbb{C}^{N_G \times L_{BR}}$. Instead of learning a mapping from a noisy image directly to a denoised one, learning the residual noise is more beneficial. Therefore, the loss function used for training is given by
\begin{equation}
\mathcal{L}_1(\bm{\varpi}_1) = \frac{1}{2I_1}\sum_{i=1}^{I_1}{\|\mathcal{R}(\mathbf{C}_r^i; \bm{\varpi}}_1) - \left(\mathbf{C}_r^i - \mathbf{X}_N^i \right)\|_F^2,
\end{equation}
where $\|\cdot\|_F$ denotes the Frobenius norm and $\mathcal{D}_1 = \{(\mathbf{C}_r^i, \mathbf{X}_N^i)\}_{i=1}^{I_1}$ represents $I_1$ clean training pairs. Algorithm 1 summarizes the detailed offline training process of the proposed DnCNN.

\addtolength{\topmargin}{-0.05in}

{\bf Remark 1: }It is worth noting that the power drift may also exist in the common angle and distance estimation during stage~I. Due to the inherent column coherence{\footnote{The column coherence can be calculated by multiplying any two columns of the dictionary matrix. If the product is non-zero, we refer to these two columns as being coherent.}} of the dictionary matrix, the single power peak corresponding to the significant path may drift to different grid indices, which is referred to as the ``power drift" effect. According to \textit{Eq.~(15) (i.e., $r_s = \frac{1}{s}Z_{\triangle}(1-\theta^2)$)} in \cite{XLM-1}, the sampling points versus distance are sufficient in the case of $s = 1$. Given that in conventional systems, the number of BS antennas $N$ is not large, the power drift can be neglected in the steering matrix associated with the BS. Therefore, we only focus our attention on the grid mismatch in stage~I. However, the power drift effect becomes significant for the cascaded steering matrix at the XL-RIS, and the amplitude of the power drift may degrade the estimation accuracy of the complex gain in stage II, as it will be elaborated on later. 

\begin{algorithm}[t]
	\caption{Training for the DnCNN-based row recovery.}
	\begin{algorithmic}[1]
        \STATE {\bf Initialize} network parameters $\bm{\varpi}_1$, learning rate $l_{r,1}$;
        \FOR{episodes = 1,...,$N_{ep}$}
        \STATE Generate mini-batch training samples $\{(\mathbf{C}_r^i, \mathbf{X}_N^i)\}_{i=1}^{I_1}$;
        \STATE Set $\mathcal{L}_1(\bm{\varpi}_1) = 0$;
        \FOR{$i=1,\dots,I_1$}
        \STATE $\mathbf{c}_r^i = \frac{1}{\tau}\sum_{j=1}^{\tau}\left[\mathbf{F}_N^H \mathbf{Y}^i\right]_{:,j}, \mathbf{C}_r^i = \left[\mathbf{c}_r^i, \dots, \mathbf{c}_r^i,\right]$;
        \STATE Update loss function $\mathcal{L}_1(\bm{\varpi}_1) = \mathcal{L}_1(\bm{\varpi}_1) + \frac{1}{2I_1}\|\mathcal{R}(\mathbf{C}_r^i; \bm{\varpi}_1) - \left(\mathbf{C}_r^i - \mathbf{X}_N^i\right) \|_F^2$;
        \ENDFOR
        \STATE Calculate the gradient $\nabla\mathcal{L}_1(\bm{\varpi}_1)$;
        \STATE Update the network $\bm{\varpi}_1 = \bm{\varpi}_1 + l_r \nabla\mathcal{L}_1(\bm{\varpi}_1)$;
        \ENDFOR
        \RETURN$\bm{\varpi}_1$.
    \end{algorithmic}
\end{algorithm}

\subsection{Stage II: Individual Steering Matrix and Complex Gain Estimation}

Based on the orthogonality of the steering vector of different angles and distances, we have $\hat{\mathbf{A}}_N^H \mathbf{A}_N(\bm{\theta}, \bm{r}) \approx \mathbf{I}_{L_{BR}}$. Hence, the received signal matrix $\mathbf{Y}$ can be projected onto the common arrival steering matrix subspace as
\begin{equation}\label{trans_Y}
\frac{1}{\sqrt{p}}\hat{\mathbf{A}}_N^H\mathbf{Y} \approx \text{diag}(\bm{\rho}) \mathbf{A}_M^H(\bm{\phi}, \bm{s}) \text{diag}(\mathbf{h}) \mathbf{E} + \frac{1}{\sqrt{p}}\hat{\mathbf{A}}_N^H\mathbf{N}.
\end{equation}
This subsection aims for estimating the cascaded matrix $\mathbf{H}_{RIS} = [\mathbf{h}_{RIS,1},\dots, \mathbf{h}_{RIS,L_{BR}}]^H = \text{diag}(\bm{\rho}) \mathbf{A}_M^H(\bm{\phi}, \bm{s}) \text{diag}(\mathbf{h})$ in (\ref{trans_Y}) for each user, which contains the individual angles, distances and complex gains. Therefore, we transform the estimation of each row of $\mathbf{H}_{RIS}$ into a sparse signal recovery problem. In particular, we define
\begin{equation}
\frac{1}{\sqrt{p}}\hat{\mathbf{A}}_N^H\mathbf{Y} \triangleq \left[\mathbf{p}_1, \dots, \mathbf{p}_{L_{BR}}\right]^H,
\end{equation}
\begin{equation}
\frac{1}{\sqrt{p}}\hat{\mathbf{A}}_N^H\mathbf{N} \triangleq \left[\mathbf{n}_{noise,1}, \dots, \mathbf{n}_{noise,L_{BR}}\right]^H.
\end{equation}
Furthermore, we have
\begin{equation}\label{pl_cs}
\begin{aligned}
\mathbf{p}_l &= \mathbf{E}^H \text{diag}(\mathbf{h}^*)\mathbf{a}_M(\phi_l, s_l) \rho_l^* + \mathbf{n}_{noise,l} \\
& = \mathbf{E}^H \text{diag}\left(\mathbf{a}_M(\phi_l,s_l)\right) [\mathbf{a}_M^*(\varphi_1, d_1),\dots, \\
& ~~~~~~~~~~~\mathbf{a}_M^*(\varphi_{L_{RU}}, d_{L_{RU}})]\bm{\beta}^*\rho_l^* + \mathbf{n}_{noise,l} \\
& = \mathbf{E}^H [\overline{\mathbf{a}}_M(\phi_l, \varphi_1, s_l, d_1), \dots, \\
& ~~~~~~~~~~~\overline{\mathbf{a}}_M(\phi_l, \varphi_{L_{RU}}, s_l, d_{L{RU}})] \bm{\beta}^*\rho_l^* + \mathbf{n}_{noise,l},
\end{aligned}
\end{equation}
where
\begin{equation}
\begin{aligned}\label{new_sparsity}
\overline{\mathbf{a}}_M(\phi_l, \varphi_p, s_l, d_p) &= \text{diag}(\mathbf{a}_M(\phi_l, s_l)) \mathbf{a}_M^*(\varphi_p, d_p) \\
& = (\mathbf{F}_M \bm{\mu}_l) \circ (\mathbf{F}_M^* \bm{v}_p^*) \\
& = (\mathbf{F}_M \circ \mathbf{F}_M^*)(\bm{\mu}_l \otimes \bm{v}_p^*) \\
& = \tilde{\mathbf{F}}_M \tilde{\mathbf{x}}_p,
\end{aligned}
\end{equation}
In (\ref{new_sparsity}), $\bm{\mu}_l \in \mathbb{C}^{M_G \times 1}$ and $\bm{v}_p \in \mathbb{C}^{M_G \times 1}$ represent the sparse vectors in the polar domain, $\tilde{\mathbf{F}}_M \in \mathbb{C}^{M \times M_G^\prime}$ is a combination of $M_G^\prime$ distinct column vectors of $\mathbf{F}_M \circ \mathbf{F}_M^*$, $\tilde{\mathbf{x}}_p \in \mathbb{C}^{M_G^\prime \times 1}$ is a sparse vector related to $\tilde{\mathbf{F}}_M$ and has only a nonzero element, which avoids the gridding complexity of $M_G^{2}$.

\begin{figure}[t]
	\centerline{\includegraphics[width=0.45\textwidth]{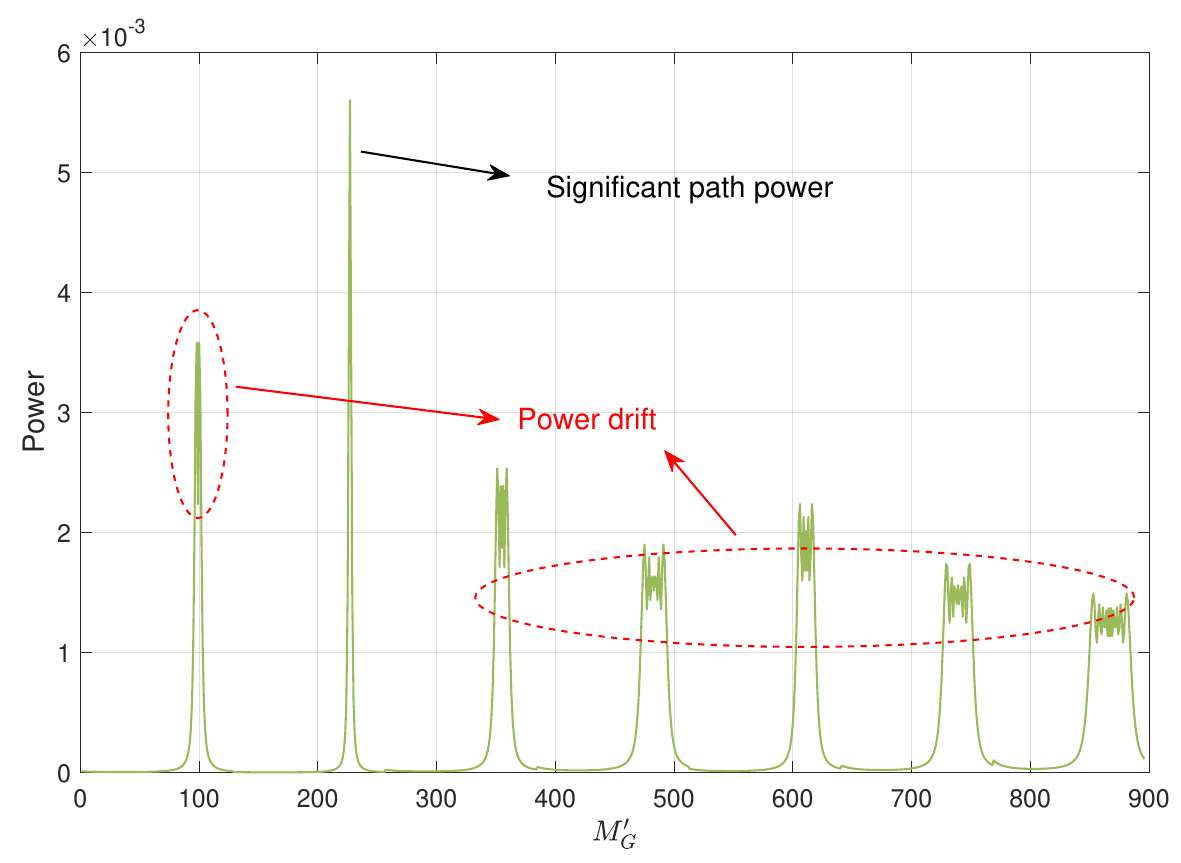}}
	\caption{Sparsity of $\mathbf{h}_{RIS,l}$, where $L_{RU}=1$ and $M=128$.}
	\label{power_leakage_2}
\end{figure}
\addtolength{\topmargin}{-0.05in}

Accordingly, (\ref{pl_cs}) can be written as 
\begin{equation}\label{pl_cs_3}
\mathbf{p}_l = \mathbf{E}^H \tilde{\mathbf{F}}_M \mathbf{b}_l + \mathbf{n}_{noise,l},
\end{equation}
where $\mathbf{b}_l \in \mathbb{C}^{M_G^\prime \times 1}$ is a sparse vector containing $L_{RU}$ nonzero elements corresponding to the cascaded complex gains $\{\beta_p^*\rho_l^*\}_{p=1}^{L_{RU}}$. In this case, (\ref{pl_cs_3}) represents a typical CS problem, where $\mathbf{E}^H$ is the measurement matrix and $\tilde{\mathbf{F}}_M$ is the dictionary matrix. Next, $\mathbf{H}_{RIS}$ can be recovered by $\hat{\mathbf{H}}_{RIS} = [\tilde{\mathbf{F}}_M \hat{\mathbf{b}}_1, \dots, \tilde{\mathbf{F}}_M \hat{\mathbf{b}}_{L_{BR}}]^H$.

However, in addition to the mismatch between the discrete estimated angles (or distances) and real continuous angles (or distances), there is a more significant challenge in the representation of the cascaded channel in the polar domain, which is the power drift. Specifically, as shown in Fig.~\ref{power_leakage_2}, the single power peak corresponding to the original significant path is drifted to different grid indices, which is caused by the coherence between the column vectors of the dictionary matrix. On the one hand, the transposed Khatri-Rao product in the dictionary matrix increases the number of sidelobes. On the other hand, the mismatch in the distance aggregates the degree of drift power imposed by the nonlinear mapping of the index to the distance sampling. Although a band-pass filter may be able to pick out the original single peak, it fails to capture the amplitude information of the genuine significant path and is incapable of correcting grid mismatches due to its overly simplistic philosophy. In this case, the traditional off-grid solutions in \cite{RIS-CE-1,XLM-1} may not be applicable to the cascaded near-field channel. To tackle these issues, we propose an ISTA-based network relying on an adaptive dictionary matrix.

\begin{algorithm}[t]
	\caption{Training for the ISTA-based network with an adaptive dictionary matrix.}
	\begin{algorithmic}[1]
		\STATE {\bf Initialize} network parameters $\bm{\varpi}_2$, learning rate $l_{r,2}$;
		\FOR{episodes = 1,...,$N_{ep}$}
		\STATE Generate mini-batch training pairs;
		\STATE Set $\mathcal{L}_2(\bm{\varpi}_2) = 0$;
		\FOR{$i=1,\dots,I_2$}
		\FOR{$l=1,\dots,L_{BR}$}
		\STATE $\mathbf{p}_l^i = [\frac{1}{\sqrt{p}}\hat{\mathbf{A}}_N^H \mathbf{Y}^i]_{:,l}^H$;
		\STATE Update loss function $\mathcal{L}_2(\bm{\varpi}_2) = \mathcal{L}_2(\bm{\varpi}_2) + \frac{1}{2I_2 L_{BR}}\frac{\|\mathcal{N}(\mathbf{p}_l^i; \bm{\varpi}_2, \tilde{\mathbf{F}}) - \mathbf{h}_l^i\|_2^2}{\|\mathbf{h}_l^i\|_2^2}$;
		\ENDFOR
		\ENDFOR
		\STATE Calculate the gradient $\nabla\mathcal{L}_2(\bm{\varpi}_2)$;
		\STATE Update the network $\bm{\varpi}_2 = \bm{\varpi}_2 + l_r \nabla\mathcal{L}_2(\bm{\varpi}_2)$;
		\ENDFOR
		\RETURN$\bm{\varpi}_2$, $\tilde{\mathbf{F}}$.
	\end{algorithmic}
\end{algorithm}

The pivotal idea of the ISTA is to recover the sparse vector $\mathbf{b}_l$ by using the following recursive equation
\begin{equation}\label{pl_cs_2}
\mathbf{b}_l^{(t+1)} = \eta_{(\lambda)}\left(\mathbf{b}_l^{(t)} - \kappa \mathbf{\Psi}^H \left(\mathbf{\Psi} \mathbf{b}_l^{(t)} - \mathbf{p}_l\right)\right),
\end{equation}
where $\mathbf{\Psi} \in \mathbb{C}^{\tau \times M_G^\prime}$ is the sensing matrix, $\mathbf{b}_l^{(t)}$ and $\mathbf{b}_l^{(t+1)}$ represent the input and output of the $t$th iteration, $\kappa$ is the iterative step size, while $\eta_{(\lambda)}(x) = \text{sign}(x)\left(|x|-\lambda\right)$ is a soft threshold function having the threshold parameter $\lambda$. Furthermore,  we treat the dictionary matrix as a ``learnable" network parameter $\tilde{\mathbf{F}}$ for adaptive adjustment. The $t$th layer of the network can be formulated as
\begin{align}
\mathbf{h}_l^{(t+1)} =\tilde{\mathbf{F}} \cdot \eta_{(\lambda^{(t)})}\left(\tilde{\mathbf{F}}^H\left(\mathbf{h}_l^{(t)} - \kappa^{(t)} \mathbf{V}\left(\mathbf{E}^H \mathbf{h}_l^{(t)} - \mathbf{p}_l\right)\right)\right),
\end{align}
where $\mathbf{h}_l^{(t)}$, $\mathbf{h}_l^{(t+1)}$ represent the input and the output of the $t$th layer, $\lambda^{(t)}$ and $\kappa^{(t)}$ represent the ``learnable" parameters of the $t$th layer, while $\mathbf{V} \in \mathbb{C}^{M \times \tau}$ represents the ``learnable" parameter shared by all layers. Let us define $\bm{\varpi}_2 = \left\{\lambda^{(t)}, \kappa^{(t)}, \mathbf{V}\right\}$. The training pairs set is represented by $\mathcal{D}_2 = \left\{[\mathbf{p}_l^i, \mathbf{h}_l^i]| \mathbf{p}_l^i = \mathbf{E}^H \mathbf{h}_l^i + \mathbf{n}, l = 1,\dots, L_{BR}\right\}_{i=1}^{I_2}$. Therefore, the loss function is given by

\begin{equation}
\mathcal{L}_2(\bm{\varpi}_2, \tilde{\mathbf{F}}) = \frac{1}{2I_2 L_{BR}} \sum_{i=1}^{I_2} \sum_{l=1}^{L_{BR}} \frac{\|\mathcal{N}\left(\mathbf{p}_l^i;\bm{\varpi}_2, \tilde{\mathbf{F}}\right) - \mathbf{h}_l^i\|_2^2}{\|\mathbf{h}_l^i\|_2^2},
\end{equation}
where $\mathcal{N}\left(\mathbf{p}_l^i;\bm{\varpi}_2, \tilde{\mathbf{F}}\right)$ is the output of the proposed network.

The network is trained offline as shown in Algorithm 2 and the trained network may then be used online at a low complexity. To elaborate, $\mathbf{h}_l^{(0)}$ is initialized by $\mathbf{h}_l^{(0)} = \mathbf{0}$, while $\tilde{\mathbf{F}}$ is initialized by $\tilde{\mathbf{F}} = \tilde{\mathbf{F}}_M$ for faster convergence. Additionally, the learnable parameters avoid the manual adjustment of hyperparameters such as $\lambda$ and $\kappa$. The adaptive dictionary matrix reduces the coherence between the column vectors during the training.

{\bf Remark 2: }As a result of the grid mismatch, the power leakage continues to be evident in the process of recovering the distances and angles at the RIS as well as the complex gains associated with each user in stage II. Fortunately, this issue would be addressed by the learned dictionary matrix and we focus only on the power drift in stage II.

\subsection{Complexity Analysis}

This subsection firstly analyzes the computational complexity of the proposed two-stage channel estimation scheme. Due to the offline nature of the proposed network's training, we examine only the complexity of online prediction. 
Specifically, the complexity order of the DnCNN is $\mathcal{O}\left(N_G L_{BR} \sum_{l=1}^{L_{c}} \left(D_x^{(l)}D_y^{(l)} D_z^{(l)} c^{(l-1)} c^{(l)}\right)\right)$ \cite{dncnn_cpx}, where $N_G$ and $L_{BR}$ correspond to the output tensor dimension, $L_c$ is the number of the DnCNN layers, $D_x^{(l)}$, $D_y^{(l)}$ and $D_z^{(l)}$ represent the tensor dimension of the convolutional kernel in the $l$th layer, and finally, $c^{(l)}$ represents the number of convolution kernels in the $l$th layer. In stage~II, the complexity order of the ISTA network is $\mathcal{O}\left(2\tau ML_{BR}L_d + ML_{BR}L_d + 2MM_G L_{BR} L_d\right)$, where $L_d$ denotes the number of layers. Since the computational complexity of the XL-RIS system is dependent on $N$ and $M$, the proposed algorithm achieves reduced complexity compared to the OMP algorithm with the complexity order of $\mathcal{O}\left(\tau N M_G N_G L_{BR} L_{RU}\right)$ \cite{RIS-CE-1}.

Additionally, the gridding complexity order of the two stages is $\mathcal{O}(N_G)$ and $\mathcal{O}(M_G)$, respectively. The two-stage approach reduces the gridding complexity order from $\mathcal{O}(N_G M_G)$ to $\mathcal{O}(N_G + M_G)$. 
Accordingly, the input layer dimension of the DnCNN is proportional to $N_G$ and the hidden layer dimension of the ISTA network is proportional to $M_G$. Thus, we have an efficient channel estimation process for the XL-RIS regime of interest.

\section{Simulation results}

\begin{figure}[t]
	\centerline{\includegraphics[width=0.5\textwidth]{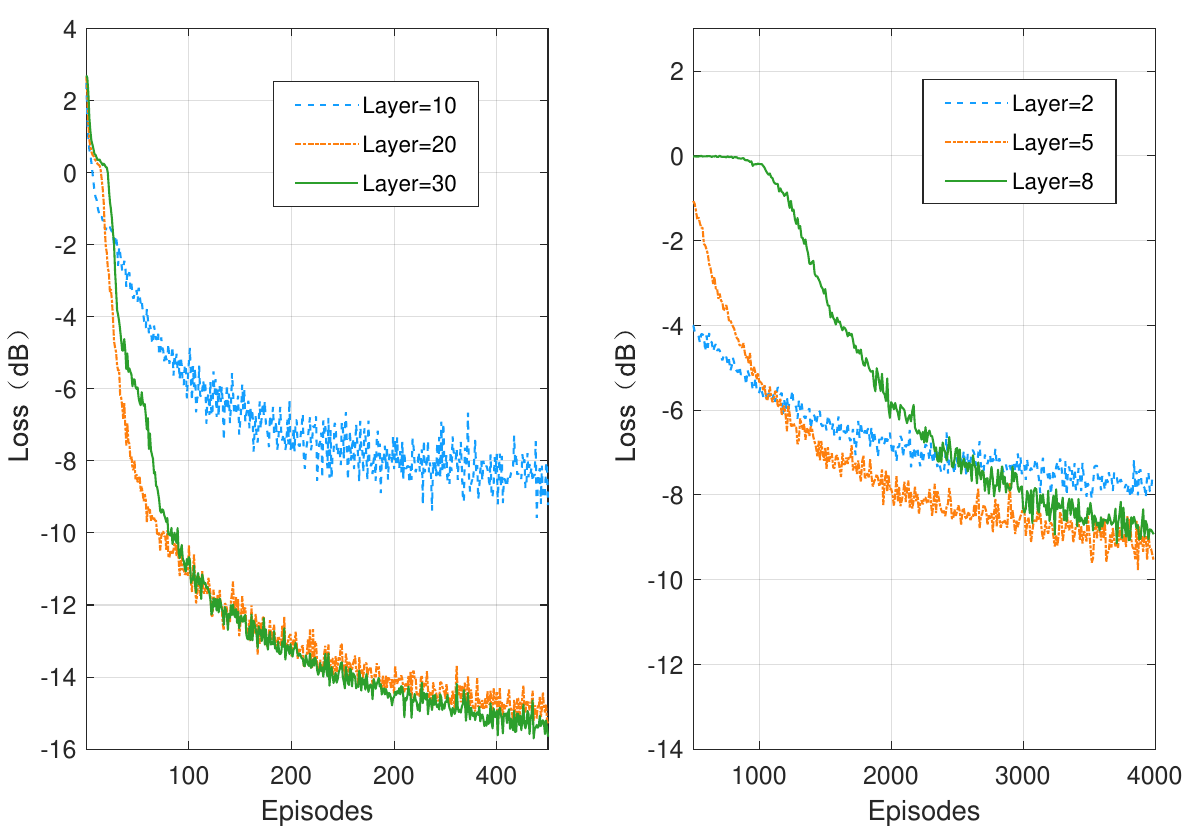}}
	\caption{The loss function of two trained networks with the learning rate of $5 \times 10^{-5}$ and $1 \times 10^{-4}$ respectively.}
	\label{result_1_loss}
	\vspace{-3mm}
\end{figure}

In this section, simulation results are provided for the performance evaluation of the proposed scheme. As for the simulation parameters, we set $f_c = 30$ GHz, $N = 64$, $M = 128$, $L_{BR} = 3$, $L_{RU} = 3$, $l_{r,1} = 5 \times 10^{-5}$, $l_{r,2} = 1 \times 10^{-4}$. Fig.~\ref{result_1_loss} plots the loss function of the two networks for different numbers of layers. Upon increasing the number of training episodes, both networks tend to converge. Increasing the  number of layers is conducive to the improvement of system performance, but the benefits obtained gradually saturate when the number of layers is increased. It is observed in Fig.~\ref{result_1_loss} that the performance of the ISTA-based network is slightly degraded when the number of layers increases to 8. This is caused by using insufficient network training.

Furthermore, two benchmarks are adopted for comparison with our proposed scheme. (i) \textbf{OMP}: The OMP algorithm is employed to solve problem (\ref{received}) based on the sparsity in (\ref{G_sparse}), resulting in the gridding complexity order of $\mathcal{O} \left( N_G M_G \right) $. (ii) \textbf{DnCNN-OMP}: The common AoAs and distances at the BS are recovered by the proposed DnCNN in stage~I, while the individual angles and distances at the XL-RIS are estimated by OMP in stage~II. (iii) \textbf{DnCNN-ISTANET}: Our proposed near-field cascaded channel estimation for the XL-RIS regime.

\begin{figure}[t]
	\centerline{\includegraphics[width=0.45\textwidth]{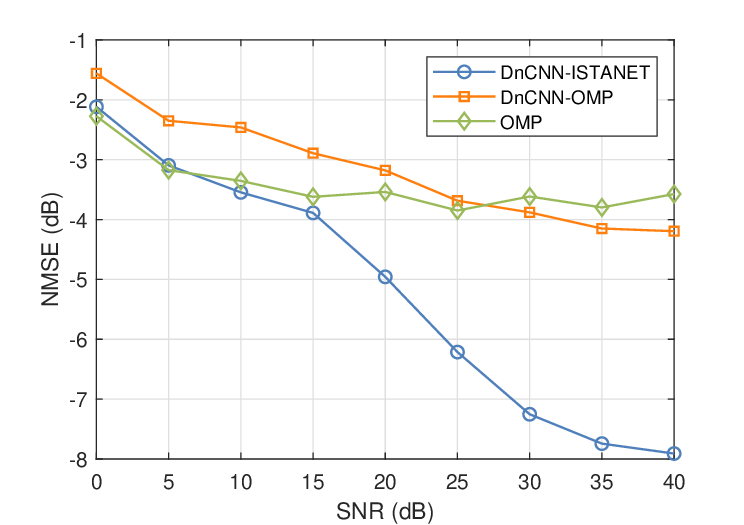}}
	\caption{NMSE versus the SNR, when $\tau=30$.}
	\label{result_2_snr}
	\vspace{-5mm}
\end{figure}

Fig.~\ref{result_2_snr} quantifies the normalized mean square error (NMSE) performance (which is defined by $\mathbb{E} \left\lbrace \frac{\|\hat{\mathbf{G}}-\mathbf{G}\|_F^2}{\|\mathbf{G}\|_F^2} \right\rbrace$) versus the signal-to-noise ratio (SNR). Observe that the NMSE decreases, when the SNR increases. Although the performance of OMP is better than that of DnCNN-OMP at low SNRs, the OMP scheme entails excessive estimation overhead for XL-RIS systems. As the SNR increases, the curve of the OMP scheme no longer decays beyond 15~dB, which is attributed to the power leakage. By contrast, the NMSE of the DnCNN-OMP scheme improves, implying that the proposed DnCNN successfully mitigates the power leakage. The higher the SNR, the better the DnCNN performs. Furthermore, the DnCNN-ISTANET scheme outperforms other benchmarks, regardless of the SNR.

\begin{figure}[t]
\centerline{\includegraphics[width=0.45\textwidth]{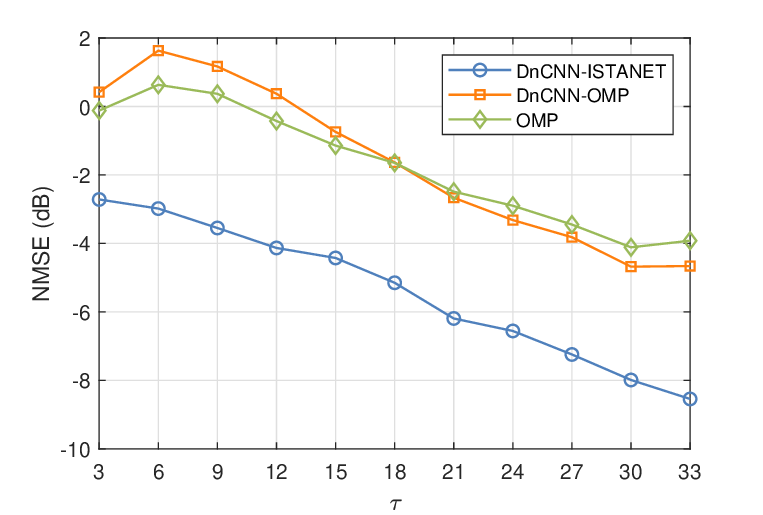}}
\caption{NMSE versus the pilot length $\tau$, when SNR $=40 ~\text{dB}$.}
\label{result_3_tau}
\vspace{-5mm}
\end{figure}

Fig.~\ref{result_3_tau} compares the NMSE performance versus the pilot length of different schemes. As expected, the increase in the pilot length improves the estimation accuracy and the proposed scheme performs better than the pair of benchmarks. In addition, the performance of the DnCNN-OMP scheme gradually exceeds that of the OMP scheme at $\tau = 18$. Accordingly, the DnCNN is more sensitive to the pilot length.
\addtolength{\topmargin}{-0.05in}

\section{Conclusion}
The cascaded channel estimation of XL-RIS aided wireless communications was investigated. Given the spherical wavefront of near-field communications, its polar-domain representation was employed to capture the potential sparsity of the near-field cascaded channels. We demonstrated that the estimation performance suffers, when using conventional CS-based approaches, which is a consequence of the associated  power leakage and power drift. To tackle these issues, inspired by the dual-structured sparsity of the cascaded channel matrix, we proposed a two-stage scheme for efficient channel estimation using learning-based approaches.
Finally, our simulation results demonstrated that the proposed learning-based two-stage scheme achieves better NMSE performance than the benchmark schemes adopted. 


\section*{Acknowledgment}
This work was supported in part by the Beijing Natural Science Foundation under Grant~4222011. Q. Wu's work is supported by NSFC 62371289 and FDCT under Grant 0119/2020/A3. The work of M. Di Renzo was supported in part by the European Commission through the H2020 ARIADNE project under grant agreement number 871464 and through the H2020 RISE-6G project under grant agreement number 101017011, and by the Agence Nationale de la Recherche (France 2030, ANR PEPR Future Networks, grant NF-PERSEUS, 22-PEFT-004). L. Hanzo would like to acknowledge the financial support of the Engineering and Physical Sciences Research Council projects EP/W016605/1, EP/X01228X/1 and EP/Y026721/1 as well as of the European Research Council's Advanced Fellow Grant QuantCom (Grant No. 789028).

\bibliographystyle{IEEEtran}
\bibliography{gc_w}

\begin{thebibliography}{10}
\providecommand{\url}[1]{#1}
\csname url@samestyle\endcsname
\providecommand{\newblock}{\relax}
\providecommand{\bibinfo}[2]{#2}
\providecommand{\BIBentrySTDinterwordspacing}{\spaceskip=0pt\relax}
\providecommand{\BIBentryALTinterwordstretchfactor}{4}
\providecommand{\BIBentryALTinterwordspacing}{\spaceskip=\fontdimen2\font plus
\BIBentryALTinterwordstretchfactor\fontdimen3\font minus
  \fontdimen4\font\relax}
\providecommand{\BIBforeignlanguage}[2]{{%
\expandafter\ifx\csname l@#1\endcsname\relax
\typeout{** WARNING: IEEEtran.bst: No hyphenation pattern has been}%
\typeout{** loaded for the language `#1'. Using the pattern for}%
\typeout{** the default language instead.}%
\else
\language=\csname l@#1\endcsname
\fi
#2}}
\providecommand{\BIBdecl}{\relax}
\BIBdecl

\bibitem{Rose}
W.~Wu, F.~Yang, F.~Zhou, Q.~Wu, and R.~Q. Hu, ``Intelligent resource allocation
  for {IRS}-enhanced {OFDM} communication systems: A hybrid deep reinforcement
  learning approach,'' \emph{IEEE Trans. Wireless Commun.}, vol.~22, no.~6, pp.
  4028--4042, Jun. 2023.

\bibitem{Ruiqi}
R.~Liu, Q.~Wu, M.~Di~Renzo, and Y.~Yuan, ``A path to smart radio environments:
  An industrial viewpoint on reconfigurable intelligent surfaces,'' \emph{IEEE
  Wireless Commun.}, vol.~29, no.~1, pp. 202--208, Feb. 2022.

\bibitem{chen-vtm}
Y.~Chen, Y.~Wang, J.~Zhang, P.~Zhang, and L.~Hanzo, ``Reconfigurable
  intelligent surface ({RIS})-aided vehicular networks: Their protocols,
  resource allocation, and performance,'' \emph{IEEE Veh. Technol. Mag.},
  vol.~17, no.~2, pp. 26--36, Jun. 2022.

\bibitem{RIS-CE-1}
G.~Zhou, C.~Pan, H.~Ren, P.~Popovski, and A.~L. Swindlehurst, ``Channel
  estimation for {RIS}-aided multiuser millimeter-wave systems,'' \emph{IEEE
  Trans. Signal Process.}, vol.~70, pp. 1478--1492, Mar. 2022.

\bibitem{RIS-CE-3}
X.~Wei, D.~Shen, and L.~Dai, ``Channel estimation for {RIS} assisted wireless
  communications—{Part II}: An improved solution based on double-structured
  sparsity,'' \emph{IEEE Commun. Lett.}, vol.~25, no.~5, pp. 1403--1407, May
  2021.

\bibitem{chen-twc3}
Y.~Chen, Y.~Wang, and Z.~Wang, ``Reconfigurable intelligent surface aided
  high-mobility millimeter wave communications with dynamic dual-structured
  sparsity,'' \emph{IEEE Trans. Wireless Commun.}, vol.~22, no.~7, pp.
  4580--4599, Jul. 2023.

\bibitem{guo-tcom}
X.~Guo, Y.~Chen, and Y.~Wang, ``Wireless beacon enabled hybrid sparse channel
  estimation for {RIS}-aided mmwave communications,'' \emph{IEEE Trans.
  Commun.}, vol.~71, no.~5, pp. 3144--3160, May 2023.

\bibitem{JSTSP}
Y.~Han, S.~Jin, C.-K. Wen, and T.~Q.~S. Quek, ``Localization and channel
  reconstruction for extra large {RIS}-assisted massive {MIMO} systems,''
  \emph{IEEE J. Sel. Topics Signal Process.}, vol.~16, no.~5, pp. 1011--1025,
  Aug. 2022.

\bibitem{tvt}
S.~Yang, W.~Lyu, Z.~Hu, Z.~Zhang, and C.~Yuen, ``Channel estimation for
  near-field {XL-RIS}-aided mmwave hybrid beamforming architectures,''
  \emph{IEEE Trans. Veh. Technol.}, to appear, 2023.

\bibitem{XLM-1}
M.~Cui and L.~Dai, ``Channel estimation for extremely large-scale {MIMO}:
  Far-field or near-field?'' \emph{IEEE Trans. Commun.}, vol.~70, no.~4, pp.
  2663--2677, Apr. 2022.

\bibitem{near_0}
O.~Rinchi, A.~Elzanaty, and M.-S. Alouini, ``Compressive near-field
  localization for multipath {RIS}-aided environments,'' \emph{IEEE Commun.
  Lett.}, vol.~26, no.~6, pp. 1268--1272, Jun. 2022.

\bibitem{near_1}
B.~Friedlander, ``Localization of signals in the near-field of an antenna
  array,'' \emph{IEEE Trans. Signal Process.}, vol.~67, no.~15, pp. 3885--3893,
  2019.

\bibitem{learning-1}
Y.~Wang, X.~Chen, H.~Yin, and W.~Wang, ``Learnable sparse transformation-based
  massive {MIMO} {CSI} recovery network,'' \emph{IEEE Commun. Lett.}, vol.~24,
  no.~7, pp. 1468--1471, 2020.

\bibitem{JiangXie}
H.~Wang, B.~Kim, J.~Xie, and Z.~Han, ``How is energy consumed in smartphone
  deep learning apps? {Executing} locally vs. remotely,'' in \emph{Proc. of
  IEEE GLOBECOM}, Waikoloa, HI, USA, Dec. 2019, pp. 1--6.

\bibitem{dncnn_cpx}
A.~Abdallah, A.~Celik, M.~M. Mansour, and A.~M. Eltawil, ``{RIS}-aided mmwave
  {MIMO} channel estimation using deep learning and compressive sensing,''
  \emph{IEEE Trans. Wireless Commun.}, vol.~22, no.~5, pp. 3503--3521, May
  2023.

\end{thebibliography}

\end{document}